\documentclass[12pts,
aps,
floatfix,
letterpaper,
prx,
longbibliography,
singlecolumn,
reprint,
superscriptaddress]{revtex4-2}
\usepackage[up,bf,raggedright]{titlesec}

\usepackage[sectionbib]{bibunits}

\usepackage{newtxtext,newtxmath}

\usepackage[colorlinks=true, linkcolor=blue]{hyperref}

\usepackage{lipsum}
\usepackage{graphicx} 
\usepackage{amsfonts}
\usepackage{amsmath}
\usepackage{braket}
\usepackage{gensymb}
\usepackage{hyperref}
\usepackage{dsfont}
\usepackage[capitalize]{cleveref}

\usepackage[utf8]{inputenc}

\date{\today}

\begin{document}

\title{Massively multiplexed nanoscale magnetometry with diamond quantum sensors}
\author{Kai-Hung Cheng} 
\affiliation{Princeton University, Department of Electrical and Computer Engineering, Princeton, NJ 08544, USA}
\author{Zeeshawn Kazi} 
\affiliation{Princeton University, Department of Electrical and Computer Engineering, Princeton, NJ 08544, USA}
\author{Jared Rovny} 
\affiliation{Princeton University, Department of Electrical and Computer Engineering, Princeton, NJ 08544, USA}
\author{Bichen Zhang} 
\affiliation{Princeton University, Department of Electrical and Computer Engineering, Princeton, NJ 08544, USA}
\author{Lila S.\ Nassar} 
\affiliation{Princeton University, Department of Physics, Princeton, NJ 08544, USA}
\author{Jeff D. Thompson} 
\affiliation{Princeton University, Department of Electrical and Computer Engineering, Princeton, NJ 08544, USA}
\author{Nathalie P.\ de Leon}
\thanks{Corresponding author. Email: npdeleon@princeton.edu}
\affiliation{Princeton University, Department of Electrical and Computer Engineering, Princeton, NJ 08544, USA}

\maketitle
\onecolumngrid
\textbf{Single nitrogen vacancy (NV) centers in diamond have been used extensively for high-sensitivity nanoscale sensing, but conventional approaches use confocal microscopy to measure individual centers sequentially~\cite{taylor_high-sensitivity_2008, maze_nanoscale_2008, lovchinsky_nuclear_2016, kucsko_nanometre-scale_2013, kolkowitz_sensing_2012, dolde_electric-field_2011}, limiting throughput and access to non-local physical properties. 
Here we design and implement a multiplexed NV sensing platform that allows us to read out many single NV centers simultaneously using a low-noise camera. 
Using this platform, we coherently manipulate and read out the spin states of hundreds of individual NV centers in parallel, achieving comparable magnetic field sensitivity to confocal measurements. 
We also implement a parallelized version of spin-to-charge-conversion readout for low NV center spin state readout noise~\cite{shields_efficient_2015} and use it to demonstrate multiplexed covariance magnetometry~\cite{rovny_nanoscale_2022}, in which we measure six two-point magnetic field correlators from four NV centers simultaneously. The number of correlators we can measure is limited only by the available laser power, opening the door to massively multiplexed covariance magnetometry. In contrast to scanning probes for imaging spatially varying magnetic fields~\cite{marchiori_nanoscale_2022}, our approach measures the temporal dynamics of the magnetic field at many precisely defined positions simultaneously, from which the local noise spectrum and non-local properties such as correlation functions can be computed. This will find immediate applications to studying condensed matter phenomena that are characterized by the noise spectrum or correlation functions~\cite{rovny_new_2024}, including quantum phase transitions~\cite{machado_quantum_2023}, dynamics far from equilibrium~\cite{Xue2024,andersen_electron-phonon_2019, zhang_nanoscale_2024}, magnetic order~\cite{ziffer_quantum_2024,Li2024a}, and correlated electron phenomena~\cite{Chatterjee2019}.}

Nitrogen vacancy (NV) centers in diamond have long spin coherence times across a wide range of temperatures combined with optical spin initialization and readout, enabling broad applications in nanoscale sensing of quantities such as magnetic field, electric field, and temperature~\cite{taylor_high-sensitivity_2008, maze_nanoscale_2008, lovchinsky_nuclear_2016, kucsko_nanometre-scale_2013, kolkowitz_sensing_2012, dolde_electric-field_2011}. NV centers are typically deployed either as individual sensors in bulk diamond~\cite{taylor_high-sensitivity_2008,maze_nanoscale_2008,lovchinsky_nuclear_2016,sutula_large-scale_2023,kolkowitz_sensing_2012,Childress2006}, scanning tip geometries~\cite{Maletinsky2012,ariyaratne_nanoscale_2018}, or in dense ensembles of NV centers that are not individually resolvable~\cite{glenn_high-resolution_2018,glenn_micrometer-scale_2017,barry_optical_2016, kazi_direct_2024,rodgers_diamond_2024}.  Resolvable NV centers offer nanometer sensing volumes~\cite{Pham2016} and high sensitivity even for NV centers within a few nanometers of the surface with proper surface preparation~\cite{sangtawesin_origins_2019,yuan_charge_2020} but suffer from slow measurements due to the high readout noise of the NV center spin state~\cite{doherty_nitrogen-vacancy_2013,Barry2020}. NV center ensembles offer fast, wide-field measurements by averaging over the signal measured by many NV centers~\cite{glenn_high-resolution_2018,glenn_micrometer-scale_2017,Barry2020,kazi_wide-field_2021} at the cost of micron-scale spatial averaging imposed by the diffraction limit. 

In this work, we demonstrate a multiplexed NV sensing platform that combines the best of these modalities, realizing parallel, wide-field measurements of many individually resolvable NV centers using a low-noise camera. We first demonstrate parallel continuous-wave (CW) ODMR of hundreds of resolvable NV centers, achieving a magnetic field sensitivity within a factor of 5 of confocal measurements of single centers in the same sample. We extend this technique to spatially resolved spin measurements, performing parallel relaxometry ($T_1$ measurements) of 70 NV centers. Next, we demonstrate enhanced parallel spin readout by implementing spin-to-charge conversion (SCC) using a spatial light modulator (SLM) to focus a high-power ionization laser onto multiple NV centers simultaneously. We realize efficient ionization and multiplexed single-shot charge state readout with a fidelity of 88\%, which enables multiplexed SCC spin readout with low readout noise, around ten times the quantum projection noise limit. This architecture allows us to measure magnetic field correlations between four individually resolved NV centers in parallel. The scalability of the multiplexed SCC is limited by the ionization laser power, and we expect that these techniques can be extended to hundreds of NV centers in future work, enabling parallel nanoscale sensing of non-local properties at many length scales~\cite{rovny_new_2024}.

\begin{figure*}[ht]
	\centering
	\includegraphics[width=183mm]{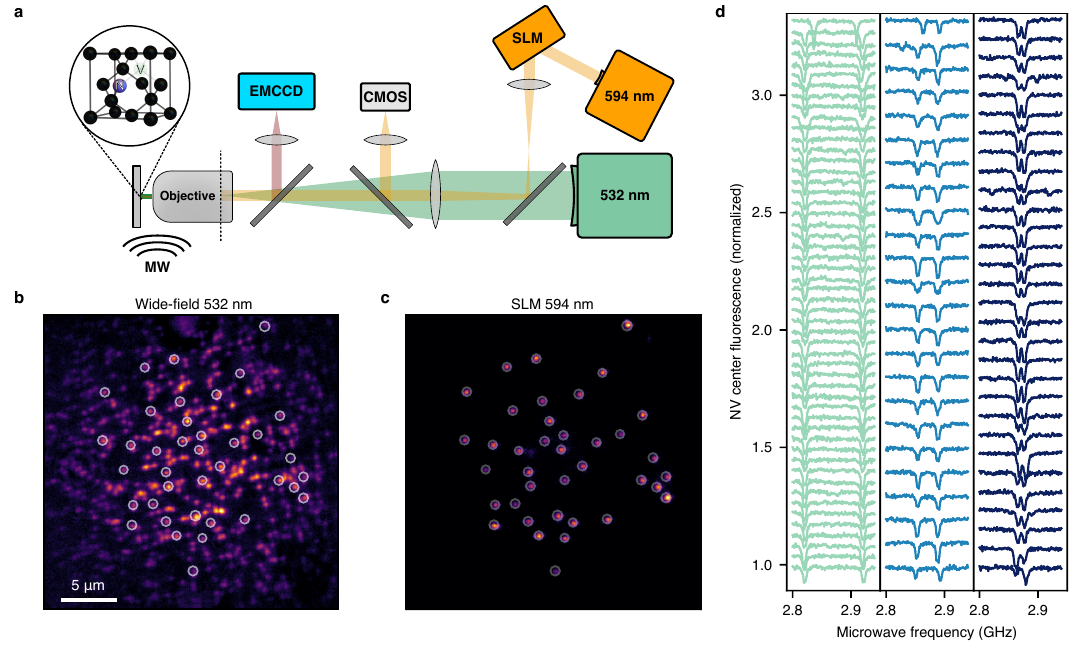}
	\caption{\textbf{Multiplexing platform for quantum sensing applications using NV centers in diamond.} 
    \textbf{a}, Diagram of the experiment setup. A green (532\;nm) laser is focused to the back focal plane of the objective lens for wide-field excitation of the NV centers. An orange (594\;nm) laser is reflected off a spatial light modulator (SLM) with a programmable phase map to generate the laser beam pattern of the selected NV centers at the sample. The SLM plane is relayed onto the back focal plane of the objective lens. The orange laser beam is sampled before the objective lens and imaged onto a CMOS camera for pattern calibration. The fluorescence emitted by the NV centers is imaged onto a low-noise electron-multiplying charge-coupled-device (EMCCD) camera for readout. (Inset) Diagram showing the NV center defect in the diamond lattice. 
    \textbf{b}, A wide-field green laser spot is used for imaging and individual spin measurement using green excitation. \textbf{c}, An orange laser beam reflected off an SLM generates a pattern of diffraction-limited laser beams at the sample to selectively excite targeted NV centers for high-fidelity readout. The circled NV centers in both images are the same.
    \textbf{d}, CW ODMR (150 spectra) measured in parallel with wide-field green laser and MW excitation. Different colors of the curves represent different families of crystallographic orientations of the NV centers.}
	\label{fig:overview}
\end{figure*}

\begin{figure*}[ht]
	\centering
	\includegraphics[width=183mm]{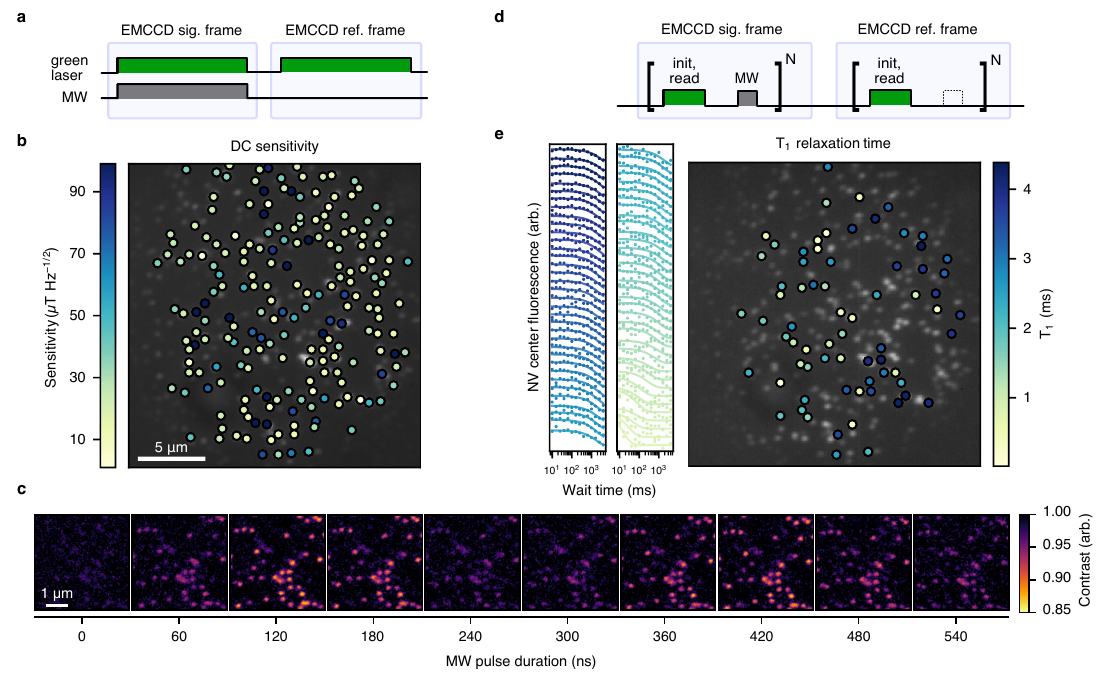}
    \caption{\textbf{Multiplexed NV spin measurements.} 
    \textbf{a}, Pulse sequence for continuous wave ODMR measurement. \textbf{b}, Map of DC magnetic field sensitivity obtained by fitting ODMR spectra and measuring photon count rates for 199 NV centers in parallel. The spatial extent of the map is given by the size of the wide-field excitation spot. Variations in sensitivity are attributed to the presence of $^{13}$C nuclei or varying proximal surface inhomogeneities leading to NV charge state instability.
    \textbf{c}, Snapshots of a wide-field Rabi oscillation experiment at increasing microwave pulse time. Each image is an average of five million experiment repetitions across two hundred camera frames.
    \textbf{d}, Pulse sequence for pulsed experiments with EMCCD camera readout. \textbf{e}, Simultaneous longitudinal spin relaxation ($T_1$) measurement of the 70 NV centers of a given crystallographic orientation (left), with a map of the fitted longitudinal spin relaxation time $T_1$ (right).
    }
	\label{fig:spinmeasurements}
\end{figure*}

Our multiplexing platform (\cref{fig:overview}\textbf{a}) uses an electron-multiplying charge-coupled device (EMCCD) camera to detect NV center fluorescence and a microwave printed-circuit board in front of the objective to deliver microwave excitation~\cite{yuan_instructional_2024}. Using CW wide-field green (532\;nm) excitation, we image several hundred NV centers with approximately 2 seconds of integration time on the camera (\cref{fig:overview}\textbf{b}). We demonstrate parallel spin control and readout by measuring the optically-detected magnetic resonance (ODMR) spectrum of all the NV centers in the field of view. We simultaneously measure several hundred ODMR spectra in parallel (\cref{fig:overview}\textbf{d}) by recording the fluorescence emitted by each NV center under CW excitation as a function of applied microwave frequency and normalizing by a reference measurement with no applied microwaves (\cref{fig:spinmeasurements}\textbf{a}). NV centers exist in four possible crystallographic orientations~\cite{taylor_high-sensitivity_2008}, and three distinct orientations are observed here (\cref{fig:overview}\textbf{d}, different colors) because two are degenerate under the static magnetic field alignment. From the ODMR spectra, we calculate the DC magnetic field sensitivity of each individual NV center and find the median DC sensitivity is 24\;\textmu T\;Hz$^{-1/2}$ (\cref{fig:spinmeasurements}\textbf{b}) compared to a median sensitivity of 5.5\;\textmu T\;Hz$^{-1/2}$ obtained with confocal measurements of the same diamond sample (Extended Data~\cref{fig:SI_dc_sensitivity,fig:SI_confocal_ODMR}).

Parallel readout also enables the multiplexed measurement of NV center dynamics. Pulsed measurements are challenging because the EMCCD camera data transfer time of 10 ms is much longer than the typical readout time for NV center spin contrast of 300 ns, determined by the lifetime of the shelving state~\cite{doherty_nitrogen-vacancy_2013}. For wide-field pulsed measurements using the camera, we repeat the experiment multiple times while integrating the fluorescence on the camera (\cref{fig:spinmeasurements}\textbf{d})~\cite{kazi_wide-field_2021}. This reduces the overhead of frame transfer time and suppresses the noise from camera readout but lowers the fluorescence contrast between the NV spin states because the camera also records fluorescence during the spin reinitialization between experiments. With this approach, we are able to measure Rabi oscillations of many NV centers simultaneously (\cref{fig:spinmeasurements}\textbf{c}). We demonstrate the application of parallel pulsed experiments to measure the spin relaxation time $T_1$ of over 70 NV centers in parallel (\cref{fig:spinmeasurements}\textbf{e}). The measured $T_1$ times in this sample, which has NV centers implanted within approximately 10\;nm of the surface, range from a few hundred microseconds to over 4\;ms, consistent with past measurements of similarly prepared samples~\cite{sangtawesin_origins_2019,Bluvstein2019}. 

We next demonstrate multiplexed low noise readout of the NV center spin state using SCC, which is critical for covariance magnetometry~\cite{rovny_nanoscale_2022} as the time required to measure a magnetic field correlation signal with a given signal-to-noise ratio scales as $\sigma_\text{R}^4$ where $\sigma_\text{R}$ is defined as the single-shot spin state readout noise~\cite{Barry2020}. SCC achieves lower $\sigma_\text{R}$ compared to green readout by using a high-power laser pulse to selectively ionize the NV center based on its spin state, followed by a high-fidelity charge state readout~\cite{shields_efficient_2015}. The selective ionization step typically requires $\sim10$ mW of optical power per NV center, so implementing SCC with a wide-field laser beam would require $>$100\;W of power for a 20 \textmu m $\times$ 20 \textmu m region. Thus, to realize multiplexed SCC in a power efficient manner, we use an SLM to create a programmable pattern of laser beams that only excite targeted NV centers (\cref{fig:overview}\textbf{c}). For simplicity, we use the same illumination pattern and wavelength (orange, 594\;nm) for both ionization and charge state readout. 

\begin{figure*}[ht]
	\centering
	\includegraphics[width=120mm]{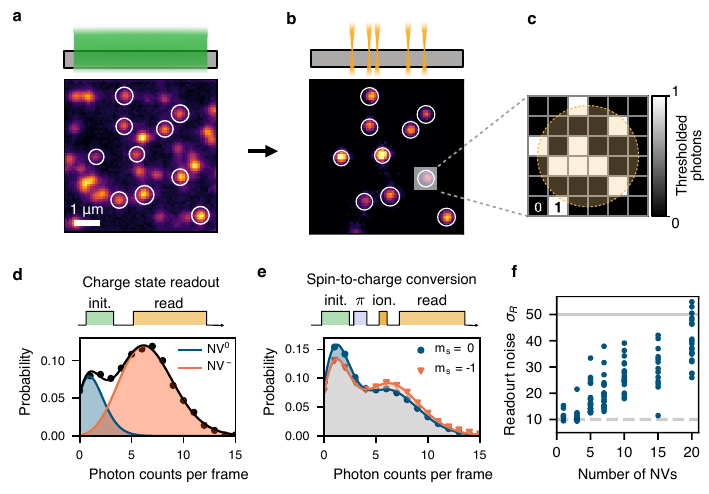}
	\caption{\textbf{Multiplexed spin-to-charge conversion readout using a spatial light modulator}. 
    \textbf{a}, Wide-field imaging using green excitation enables localization of single NV centers, whose coordinates are recorded. 
    \textbf{b}, The SLM enables selective excitation with the orange laser, enabling ionization and charge state readout. 
    \textbf{c}, Diagram of photon counting using the EMCCD camera. In a 6 pixel by 6 pixel region around each NV center (dashed circle), photons are counted by defining a detection threshold for each pixel and summing. 
    \textbf{d}, Charge state readout of an NV center using SLM excitation. (Top) Wide-field green laser pulse initializes the NV center charge state, which is then read out using a weak orange pulse whose duration is matched to the camera exposure (8\;ms). The pulse sequence is repeated several thousand times, and the resulting counts per NV center per frame are put in a histogram, with a representative histogram for one NV center shown (Bottom). The curves are fits to two Poisson distributions that correspond to the relative populations of NV$^0$ and NV$^-$, with charge state readout fidelity of $88.3 \pm 0.5 \%$. 
    \textbf{e}, Spin-to-charge conversion using the SLM. (Top). After green laser initialization, a MW $\pi$-pulse can be applied to flip the spin from $m_s=0$ to $m_s=-1$. The orange laser is then amplitude modulated to ionize the NV centers (high power, short pulse), and read out the charge state (low power, long pulse). (Bottom) The resulting photon number distributions for the $m_s=0$ and $m_s=-1$ spin states for a representative NV center when the ionization laser power is distributed over three NV centers are shown. 
    \textbf{f}, Readout noise $\sigma_\text{R}$ as a function of number of simultaneously measured NV centers. Each point represents the measured $\sigma_\text{R}$ from an NV center. At high numbers, the readout noise is limited by low ionization intensity. At low NV numbers (high ionization power) the readout noise is limited by the recombination of NV$^-$ back to NV$^0$.}
	\label{fig:SLM}
\end{figure*}

We implement multiplexed charge state readout with the SLM and the camera
through the following steps, outlined in \cref{fig:SLM}\textbf{a-c} (Methods~\cref{exptdetails}). First, we take an image of all the NV centers using wide-field green excitation and identify the NV centers that we want to target (\cref{fig:SLM}\textbf{a}). Second, we run a weighted Gerchberg-Saxton algorithm to generate a phase pattern on the SLM to create the laser spots at the fitted NV center locations (\cref{fig:SLM}\textbf{b})~\cite{nogrette_single-atom_2014}. To read out the charge state, we excite the NV centers with a low-power orange laser and measure the photon counts per exposure by thresholding individual pixels in a region around each NV center and then summing the thresholded counts~(\cref{fig:SLM}\textbf{c}). In this low-photon regime, we binarize the image such that each pixel measures either 0 or 1 photons and do not account for higher per-pixel counts, which allows us to take full advantage of the camera's readout statistics~\cite{basden_photon_2003}.
A representative histogram of photon counts per camera frame is shown in \cref{fig:SLM}\textbf{d} (Extended Data Fig.~\ref{fig:SI_chargereadout}), revealing two peaks that correspond to the NV$^0$ (dark) and NV$^-$ (bright) charge states under orange illumination. The measured charge state readout fidelity using the camera (88.3\%) is similar to that obtained using confocal readout of the same sample (86.3\%) with the same objective lens and readout parameters (Extended Data~\cref{fig:SI_confocal_charge_readout}). 

Having achieved multiplexed charge-state measurements using targeted excitation, we then add a spin-selective ionization pulse to implement multiplexed SCC by modulating the power of the orange laser between ionization ($\sim$10\;mW per spot) and readout ($\sim$20\;\textmu W per spot). We compare the charge state readout statistics for a representative NV center spin prepared in either the $m_s=0$ or $m_s=-1$ spin state to benchmark $\sigma_\text{R}$ (\cref{fig:SLM}\textbf{e}). The spin readout noise for the data shown in~\cref{fig:SLM}\textbf{e} is $\sigma_\text{R}\approx12$, a four-fold improvement over green readout in a confocal microscope~\cite{Barry2020} and a ten-fold improvement over wide-field green readout (Extended Data~\cref{fig:SI_wf_sigmaR}). 
However, our scheme performs worse than previous implementations of SCC using a red (637\;nm) laser for ionization ($\sigma_\text{R}\sim5$)~\cite{rovny_nanoscale_2022}, which we attribute to a higher charge state recombination rate from the orange laser~\cite{aslam_photo-induced_2013}, in agreement with a rate equation model~\cite{wirtitsch_exploiting_2023} (Extended Data~\cref{fig:SI_rate_eq}).

To understand the scalability of our multiplexed SCC protocol, we investigate the scaling of $\sigma_\text{R}$ as a function of the number of multiplexed NV centers with a constant total ionization pulse power (\cref{fig:SLM}\textbf{f}). We observe that $\sigma_\text{R}$ increases as we increase the number of addressed NV centers. We attribute this increase to inefficient charge state ionization, as the power available for each NV center is reduced when we distribute the fixed total laser power across more NV centers. This indicates that the scale of our multiplexed SCC experiment is only limited by the currently available ionization laser power (30 mW). 

\begin{figure*}[ht]
	\centering
	\includegraphics[width=120mm]{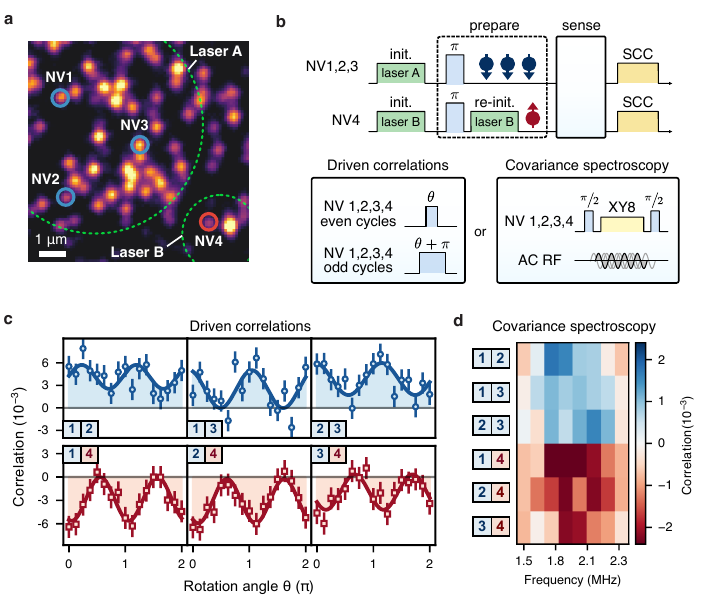}
	\caption{\textbf{Multiplexed correlation measurements}. 
    \textbf{a}, Wide-field image showing the 4 targeted NV centers. The NV center spin and charge states are initialized by lasers A and B, where A addresses NV1-3 and B addresses only NV4, enabling selective reinitialization. 
    \textbf{b}, Pulse sequences of the two types of correlation experiments carried out. We demonstrate the measurement of driven-spin correlations and spectroscopy of a correlated field using 4 NV centers. In both experiments, we prepare three NV centers (NV1-3) in the same spin state and one NV center (NV4) in the opposite spin state. In the driven-spin correlation experiment, we interleave MW pulses that rotate the spin by either $\theta$ or $\theta + \pi$ and sweep $\theta$. In the spectroscopy experiment, we run an XY8 sensing sequence with a fixed inter-pulse spacing while applying an AC magnetic field to the sample and sweeping the AC frequency.
    \textbf{c}, Driven-spin correlations between 4 NVs as a function of $\theta$. The three blue curves on the top show positive correlations between NV1-3, and the three red curves on the bottom show negative correlations between NV4 and NV1-3. The errorbars represent the standard error.
    \textbf{d}, Covariance noise spectroscopy of an applied field with 4 NV centers. The three NV pairs that start in the same spin state show positive correlations, while the three pairs that start in the opposite spin states show negative correlations. 
    }
\label{fig:multiplex_corr}
\end{figure*}

Equipped with multiplexed SCC, we demonstrate parallelized covariance magnetometry by reconstructing non-local magnetic field correlators $\braket{B(\vec{x}_i)B(\vec{x}_j)}$ from simultaneous measurements of NV centers at different locations $\vec{x}_i$~\cite{rovny_nanoscale_2022}. We target 4 NV centers (\cref{fig:multiplex_corr}\textbf{a}) from which we simultaneously measure ${4 \choose 2}=6$ two-point correlators. To demonstrate that the measured correlations result from spin dynamics rather than spurious correlations \cite{rovny_nanoscale_2022}, we use a separate green laser to prepare one of the NV centers (NV4) in the opposite spin state from the other three (NV1-3) (\cref{fig:multiplex_corr}\textbf{a,b}). To verify our ability to detect two-point correlators with our achievable readout noise, we first measure correlations introduced by direct NV spin driving rather than noise sensing (\cref{fig:multiplex_corr}\textbf{b}). This allows us to rapidly drive the NV centers to mitigate the effect of decoherence under free evolution, as well as to circumvent quantum projection noise as we are able to intentionally drive each NV center between pure states $m_s=0$ and $m_s=-1$. 
In this driven-spin correlation experiment, we interleave two pulse sequences: in the even-numbered experiments, we simultaneously drive all the NV centers around the Bloch sphere by an angle $\theta$, while in the odd-numbered experiments we drive the NV centers by an angle $\theta+\pi$ (\cref{fig:multiplex_corr}\textbf{b}). At the end of each experiment we read out a single camera frame, simultaneously measuring the four signals $s_{i}$ where $i=1...4$ indexes the NV center. Lastly, we use the set of acquired signals across experiments $S_i=\{s_{i,k}\}$, where $k$ indexes the experiment number, to compute the Pearson correlation for each NV pair: $r_{ij}=\text{Cov}(S_i,S_j)/(\sigma_i \sigma_j)$, where $\text{Cov}$ is the covariance and $\sigma_i$ is the standard deviation of $S_i$. From the computed Pearson correlations, we subtract off a uniform background correlation measured using an independent calibration experiment (Extended Data~\cref{fig:SI_bg_corr}); the resulting correlations are shown in \cref{fig:multiplex_corr}\textbf{c}. As expected for each pair of NV centers prepared in the same initial spin state (NV1-3), the correlations between the measured signals are maximized when $\theta=n \pi$ for integer $n$, as the two spins flip synchronously between $m_s$\;=\;$0$ and $-1$, and are minimized for $\theta=(n+1/2) \pi$, since the readout projects the two spins randomly and independently onto $m_s$\;=\;$0$ or $-1$. Conversely, each NV center's correlation with NV4 is negative, as NV4 has been prepared in the opposite orientation and thus acquires an opposite phase. In addition, the amplitude of the correlation oscillation indicates the maximum measurable correlation, which is $1/\sigma_{R_i}\sigma_{R_j}$. From \cref{fig:multiplex_corr}\textbf{c}, we estimate $\sigma_{R_i}=$ $15 \pm1$, consistent with the data in \cref{fig:SLM}\textbf{f}.

We also use multiplexed covariance magnetometry to perform covariance noise spectroscopy of an applied AC magnetic field~\cite{rovny_nanoscale_2022}. As before, we prepare NV1-3 in the same spin state and prepare NV4 in the opposite state. We then use an XY8 sensing sequence~\cite{Meiboom1958} with a fixed interpulse spacing, and sweep the frequency of the applied AC signal~(\cref{fig:multiplex_corr}\textbf{d}). When the frequency matches the interpulse spacing, NV1-3 show positive pair-wise correlations from the correlated phase they accumulate due to the shared noise, and NV4 shows negative pair-wise correlations with NV1-3 because of the opposite phase accumulation. 

The platform demonstrated here can be readily extended to accomplish low readout noise measurements of many more NV centers. Provided 10\;W of ionization laser power, we could address 500 NV centers resulting in more than $10^5$ measured two-point correlators. Moreover, performing ionization at a more optimal wavelength is expected to reduce $\sigma_\text{R}$ further. Finally, extending these techniques to measure correlations below the diffraction limit~\cite{Rovny2024b} would allow for the measurement of many two-point correlators with nanometer separations simultaneously.

Our platform opens the door to high-throughput measurements of magnetic fields across large areas with nanoscale localization, with immediate applications in magnetometry of condensed matter dynamics. Just as scanning NV magnetometry has advanced our understanding of magnetic ordering~\cite{Thiel2019}, transport~\cite{Ku2020,Jenkins2022, Vool2021}, and superconductivity~\cite{Thiel2016,Chen2024} through high-resolution imaging of magnetic fields, our platform will provide a new window into dynamics through wide-field mapping of local noise spectra and non-local correlation functions at many length scales simultaneously. Specifically, local noise measurements can provide a quantitative probe of dynamics~\cite{kolkowitz_probing_2015,dwyer_probing_2022}, for instance in measuring critical exponents of phase transitions in 2D antiferromagnets that have vanishing stray fields~\cite{machado_quantum_2023,ziffer_quantum_2024}. Additionally, access to non-local correlation functions with nanoscale precision will enable study of the dynamics in systems that are fundamentally characterized by correlations, such as in hydrodynamic electron flow~\cite{Agarwal2017,Lucas2018,zhang_nanoscale_2024}, magnetic 2D materials~\cite{ziffer_quantum_2024,machado_quantum_2023}, vortex dynamics in superconductors~\cite{curtis_probing_2024, chen_when_2024}, and spin liquids~\cite{Chatterjee2019,Broholm2020}.

\bibliography{nv_multiplex.bib}

\textbf{Note added:}
We are aware of a manuscript \cite{Cambria2024a} that describes complementary work to perform multiplexed measurements with NV centers in diamond. Both works deploy low-noise cameras to measure many NV centers simultaneously; in \cite{Cambria2024a} Cambria et al.\ employ a complementary approach for initialization and ionization by using an acousto-optic deflector instead of an SLM to achieve high optical power density.

\textbf{Data availability:}
The data that support the findings of this study are available from the corresponding author upon reasonable request.

\textbf{Code availability:}
The code related to the data analysis of this study is available from the corresponding author upon reasonable request.

\begin{acknowledgments}
We gratefully acknowledge helpful conversations with Pai Peng, Artur Lozovoi, Matthew Cambria, and Shimon Kolkowitz. This work was supported by the Gordon and Betty Moore Foundation (GBMF12237, DOI 10.37807), the National Science Foundation (Grant No. OMA-2326767), and the Intelligence Community Postdoctoral Research Fellowship Program by the Oak Ridge Institute for Science and Education (ORISE) through an interagency agreement between the US Department of Energy and the Office of the Director of National Intelligence (ODNI) (JR).

\end{acknowledgments}

\textbf{Author contribution:}
KHC, JR, NPdL conceived the experiment platform. KHC and ZK built the optical setup with inputs from JR, BZ, and LSN. KHC wrote the control software. BZ and JDT instructed the setup of the spatial light modulator. KHC and ZK collected and analyzed the data. KHC performed the numerical simulation and derivations. JDT and NPdL supervised the project. KHC, ZK, JR, and NPdL wrote the manuscript with input from all the authors.

\textbf{Competing interests:} The authors declare no competing interests. 

\textbf{Correspondence and requests to materials} should be addressed to N. P. de Leon.

\pagebreak
\clearpage

\widetext
\begin{center}

\end{center}

\setcounter{secnumdepth}{3}
\setcounter{equation}{0}
\setcounter{figure}{0}
\setcounter{table}{0}
\setcounter{page}{1}
\makeatletter
\renewcommand{\theequation}{S\arabic{equation}}
\setcounter{page}{1} 
\textbf{\LARGE Methods}
\section{Experimental details} \label{exptdetails}

The diamond sample was implanted with a nitrogen ion energy of 3\;keV, resulting in shallow NV centers within approximately 10\;nm of the surface. NV center measurements are performed in a home-built setup that include a wide-field illumination path, an SLM excitation branch, and an imaging path. The sample is mounted in a closed-cycle cryostat (Montana Instrument S100), optically interfaced through a Zeiss 100x, NA = 0.9 objective. The wide-field green laser used for imaging of the NV centers and wide-field spin measurements is provided by a 532\;nm diode-pumped solid state laser (Coherent Genesis STM CX-5000), modulated by an acousto-optic moodulator (AOM) (Isomet M1205-P80L-1) in a double-pass configuration. The orange laser used for spin-to-charge conversion (SCC) as well as NV center charge state readout is a 594\;nm, diode-pumped solid state laser (Hubner Cobolt Mambo) equipped with an internal AOM for both digital and analog modulation of the laser power. The two excitation lasers are combined by a dichroic mirror (Semrock AF567-Di01). The laser powers (as measured before the objective lens) used were approximately 15\;\textmu W per NV center for charge state readout, approximately 30\;mW in total for the ionization pulse of SCC (equally distributed across all the NV centers being addressed), and approximately 50\;mW for green wide-field imaging and charge state initialization. Both excitation lasers are circularly polarized before the objective (Thorlabs WPQ10M-532 and WPQ10M-588). The local re-initialization laser for demonstrating anti-correlations in covariance magnetometry is a 520\;nm diode-pumped solid state laser (Hubner Cobolt MLD-06).

The readout path is separated from the excitation by a dichroic mirror (Semrock FF647-SDi01). The NV center fluorescence is filtered to have a pass band of 650 to 800 nm and then imaged onto an EMCCD camera (Nuvu HNu-512). All the experiments were done with an EM-gain of 5000. For the wide-field spin experiments, the camera is exposed to the NV center fluorescence for multiple iterations and read out periodically. The length for each exposure is chosen empirically based on the optical intensity of the excitation laser and the laser duty cycle to prevent the camera from being over-exposed. The data shown in \cref{fig:spinmeasurements} are obtained this way. 

An SLM (Santec SLM-210) is used to generate a laser pattern on the diamond sample to selectively address the NV centers chosen for SCC readout. We incorporate an open-source SLM control software package~\cite{kim_large-scale_2019} for SLM calibration and phase pattern generation. The laser pattern is sampled from the main optical path onto a CMOS camera (Thorlabs CS165MU) for adaptive pattern calibration. The SLM pattern that corresponds to the locations of the selected NV centers is generated after selecting the NV positions, as shown in~\cref{fig:overview}\textbf{a}. 

Microwave pulses are generated using a Rohde and Schwarz signal generator (SMATE200A), amplified (Mini-Circuits ZHL-16W-43S+), and sent to a homemade microwave printed circuit board. IQ phase modulation is achieved using an arbitrary waveform generator (AWG) (Keysight 33622A) that modulates the signal generator. The low frequency AC test signal around 2 MHz is generated with another AWG (Keysight 33622A) and amplified (Mini-Circuits LZY-22+). The test signal is delivered to the sample via a copper coil loop fixed around the diamond sample.

\section{Wide-field spin measurement of NV centers}
Conventional confocal microscope NV center measurements use single-photon avalanche photodiodes that have nanosecond timing resolution and sub-microsecond readout rate. In the conventional experiment configuration, it is possible to read out each 300 ns shot of NV center experiments, which typically have repetition rates on the order of 10 kHz to MHz. However, in the wide-field case, since the EMCCD frame rate is less than 100 Hz when operating in the full frame, it is impractical to operate the camera the same way as we would for conventional NV experiments. To overcome this issue, instead of reading out every experiment shot separately, we integrate the fluorescence from multiple pulsed experiments on the EMCCD and then read out the frame after a certain number of repetitions before saturating the camera. We take a reference frame using the same approach with the same camera exposure time and laser duty cycle but without applying the MW pulse(s). The schematics of a pulse diagram is shown in \cref{fig:spinmeasurements}(\textbf{d}). This approach allows us to implement multiplexed pulsed microwave experiments using the camera, shown in~\cref{fig:spinmeasurements}\textbf{c} and~\cref{fig:spinmeasurements}\textbf{e}.

The fluorescence signal for each NV center is obtained through the following process. First, the locations of NV centers are determined from a wide-field image by running a blob-detection algorithm. After the locations of the NV centers are determined, we sum the pixel values in an n$\times$n region centered around each NV center for both signal and reference frames to obtain the signal count $c_{sig}$ and reference count $c_{ref}$. That is,
\begin{align}
    c_{sig, ref} = \sum_{i=1}^{n} \sum_{j=1}^{n} c_{sig,ref}^{raw}(i,j)
    ,\label{eq:SI_wf_PL}
\end{align}
where $c^{raw}$ denotes the raw camera reading. In this work, we choose n = 6 based on the size of NV center image on the EMCCD. The normalized fluorescence is obtained by dividing $c_{sig}$ by $c_{ref}$, i.e., $c_{norm}=c_{sig}/c_{ref}$, to cancel out slow drifts in fluorescence intensity over time.
With this data acquisition approach tailored to measuring single NV center with the conventional green readout on an EMCCD, we measure the CW-ODMR spectra of the NV centers (\cref{fig:overview}(\textbf{d})), fit the spectra, and run Rabi oscillation on our selected resonance frequency using the aforementioned pulsed experiment method to calibrate our $\pi-$pulse time for subsequent pulsed experiments.

We calculate the DC magnetic field sensitivity of the NV centers from the ODMR spectrum using
\begin{align}
    \eta \approx 
    \frac{h}{g \mu_B} \frac{\Delta \nu}{C \sqrt{\mathcal{I}_0}}, \label{eq:dc_sen}
\end{align}
where $g \approx 2$ is the Land$\acute{e}$ g-factor, $\mu_B$ is the Bohr magneton, $\Delta \nu$ is the ODMR linewidth, $C$ is the ODMR contrast, and $\mathcal{I}_0$ is the fluorescence detection rate~\cite{taylor_high-sensitivity_2008}.
We fit each individual ODMR spectrum to obtain the linewidth $\Delta \nu$ and the contrast $C$, and obtain the fluorescence rate $\mathcal{I}_0$ for each NV center from the wide-field image. The resulting DC magnetic field sensitivity map is shown in \cref{fig:spinmeasurements}\textbf{b}. Note that more than 60\% of the NV centers have a sensitivity less than 30\;\textmu T\;Hz$^{-1/2}$, comparable to the sensitivity obtained from a single NV center measured with a confocal microscope (Extended Data~\cref{fig:SI_confocal_ODMR}). The histogram of the DC magnetic field sensitivity from the wide-field dataset is shown in Extended Data Fig.~\ref{fig:SI_dc_sensitivity}.

In addition, we benchmark the spin readout noise, $\sigma_\text{R}$ using wide-field green excitation. $\sigma_\text{R}$ is defined as

\begin{align}
    \sigma_\text{R}  = \sqrt{1 + 2\frac{\sigma_0^2 + \sigma_1^2}{(\alpha_0 - \alpha_1)^2}},
    \label{eq:SI_sigmaRgen}
\end{align}
where $\sigma_{0,1}^2$ and $\alpha_{0,1}$ are the variances and means, respectively, of the $m_s=0,1$ state photon number distributions~\cite{rovny_nanoscale_2022}. The $\sigma_\text{R}$ of most NV centers are above 100, which is several times to an order of magnitude worse than the green readout using a confocal microscope (Extended Data \cref{fig:SI_wf_sigmaR}). We attribute the high readout noise to the following three causes: (1) The EMCCD, unlike confocal microscopes, does not have a pinhole in the detection optical path to block out-of-plane light, so the EMCCD readout suffers from increased background fluorescence. (2) The spin contrast we obtain from the EMCCD using the protocol described above is lower than that obtained from a confocal microscope because when we integrate multiple pulsed experiments on one EMCCD exposure, we collect the fluorescence while the NV center spin is reinitialized, reducing the spin contrast. (3) The optical power density we use is approximately two orders of magnitude lower than the saturation power of a single NV center in a confocal microscope setup.

\section{Multiplexed NV center charge state readout on the EMCCD}
To read out the NV center charge state, the NV centers are first imaged using the wide-field green laser, then the positions of the selected NV centers are  determined from the image by fitting the image to a 2D-Gaussian distribution.

The fitted NV coordinates can then be used to generate the required phase pattern on the SLM using a w-GS algorithm. During readout, the image is processed in real-time to extract the photon counts from each NV center. The photon number for a given NV center in a given readout, $S$, is obtained by integrating an n$\times$n region around the target NV center. That is 
\begin{align}
    S = \sum_{i=1}^{n} \sum_{j=1}^{n} c(i,j)
    ,\label{eq:SI_charge_state_counts}
\end{align}
where $c(i,j)$ is the photon count at pixel $(i,j)$. In this work, we choose $n=6$ based on the image size of an NV center.
To obtain the photon count at each pixel, the pixels are thresholded to determine whether they detected either 0 or 1 photon based on the raw camera readings. That is,
\begin{align}
    c(i,j) = 
    \begin{cases}
        0, & \text{if } c_{raw}(i,j) \leq t_{pc} \\
        1, & \text{if } c_{raw}(i,j) > t_{pc} \\
    \end{cases}
    ,\label{eq:SI_threshold}
\end{align}
where $c_{raw}(i,j)$ is the raw camera reading for pixel $(i,j)$ and $t_{pc}$ is the photon counting threshold.
During readout, the selected NV centers are illuminated with focused orange laser beams, and their fluorescence is collected on the EMCCD during the camera exposure. The photon-detection threshold must be large enough to surpass the readout noise of the EMCCD readout circuit. We do not observe a strong effect on both charge state population and  spin readout noise when we sweep the photon-detection threshold ($t_{pc}$) around the chosen value. The thresholded photon counts in each NV region are then summed to yield the final counts per NV center per frame. The photon counts in \cref{fig:SLM} and Extended Data \cref{fig:SI_chargereadout} are obtained in this way. As demonstration that we are able to read out the charge state of multiple NV centers in parallel, the simultaneous charge state readout histograms of 15 NV centers are plotted in Extended Data Fig.~\ref{fig:SI_chargereadout}. From the double-Poisson fit to the histogram, we extract the relative charge state populations. We attribute the low NV$^-$ population of some NV centers to surface inhomogeneities that destabilize the charge state of the near-surface NV centers.

\section{Sensitivity and charge state readout fidelity comparison to confocal microscope measurement}
To benchmark the DC magnetic field sensitivity and charge state readout fidelity of the multiplexing experiments against conventional confocal measurements, we perform CW-ODMR and charge state readout using confocal excitation and collection with the same diamond sample and objective lens. The resulting CW-ODMR spectra and the corresponding DC magnetic field sensitivities are presented in Extended Figure \cref{fig:SI_confocal_ODMR}, indicating that our wide-field sensitivity is within an order of magnitude of that measured with a confocal microscope. We attribute the discrepancy between sensitivities measured with camera and confocal readout to the lower spin state readout contrast using the camera, as described in previous sections.

A typical NV center charge state readout histogram using a confocal microscope with the same readout time as that of the histogram in \cref{fig:SLM}\textbf{d} is shown in Extended Data \cref{fig:SI_confocal_charge_readout}. The associated charge state readout fidelity is 86.3 $\pm$ 0.8$\%$, a typical value with these readout parameters, indicating that the charge state readout fidelity on the camera ($88.3 \pm 0.5 \%$ in \cref{fig:SLM}\textbf{d}) is comparable to that obtained from a confocal microscope measurement.

\section{Multiplexed spin readout noise measurement}
A key parameter for covariance magnetometry sensitivity is the spin-state readout noise $\sigma_\text{R}$. We use the SLM to implement SCC on multiple NV centers and benchmark the efficacy of the scheme by measuring the spin-dependent photon statistics for the $m_s$\;=\;0 and $m_s$\;=\;-1 spin states to obtain $\sigma_\text{R}$. Note that in order to use the same laser for both ionization and readout, we modulate the control voltage of the AOM after calibrating the input voltage and output power.
To measure the photon statistics for $m_s$\;=\;0 state, we initialize the NV center spin state and charge state and implement SCC readout. For the $m_s$\;=\;-1 state, we apply a MW $\pi-$pulse after initialization and read out using SCC. \cref{fig:SLM}\textbf{e} shows an example of the histograms of the photon counts for the two spin states for a single NV center. We then extract the statistics of the two distributions to compute $\sigma_\text{R}$ using ~\cref{eq:SI_sigmaRgen}.

Since the efficacy of SCC depends on both the ionization step that maps different spin states into different charge states and the readout step, $\sigma_\text{R}$ is a function of four parameters: ionization power, ionization time, readout power and readout time. Therefore, to optimize $\sigma_\text{R}$, we optimize these four parameters within the constraints of our experiment setup. The readout power and time are set by the sweeping these parameters and measuring the charge state populations of multiple NV centers~(Extended Data Fig.~\ref{fig:SI_chargereadout}). We use approximately 10-20\;\textmu W per NV center with a 5\;ms readout because these parameters result in a high fidelity charge state readout with reasonable experiment repetition rates. We also observe that small changes in the readout parameters have low effect on the measured $\sigma_\text{R}$. Since the ionization step is power-limited, we fix the ionization power to the maximum available (approx. 30\;mW). Therefore, to optimize $\sigma_\text{R}$, we set readout time, readout power, and total ionization power as detailed above and sweep ionization time, shown in Extended Data Fig.~\ref{fig:SI_sigmaR_tion} for 5 NV centers simultaneously. In this case, the minimum $\sigma_\text{R}$ for all 5 NV centers is achieved for ionization pulse time around 250\;ns.

\section{Limits to multiplexed readout noise}
The minimum $\sigma_\text{R}$ we have measured with our modified SCC scheme is approximately 2 to 3 times higher than observed in other works \cite{shields_efficient_2015, rovny_nanoscale_2022}. By applying a rate equation model \cite{yuan_charge_2020,wirtitsch_exploiting_2023}, we study the dependence of $\sigma_\text{R}$ on ionization laser power and wavelength. In our simulation, we assume the following parameters of the NV center initial state: 70\% NV$^-$ population, 95\% initialization into $m_s$\;=\;0 state, and 1.6/6.7 photons on average for the NV$^0$/NV$^-$ charge state readout, based on typical results of our charge state measurements. We evolve the system under the rate equation model with 95\% $m_s$\;=\;0 or $m_s$\;=\;1 initial population that correspond to the cases where we apply no MW pulse or a $\pi-$pulse. At each time step, we extract the photon statistics of both cases from their instantaneous charge state distribution, $f_0(t)$ and $f_\pi(t)$:
\begin{align}
    f_0(t) = \textrm{Poisson}(\lambda=\lambda_0) \times p^{NV^0}_{0}(t) + \textrm{Poisson}(\lambda=\lambda_1) \times p^{NV^-}_{0}(t) \\
    f_\pi(t) = \textrm{Poisson}(\lambda=\lambda_0) \times p^{NV^0}_{\pi}(t) + \textrm{Poisson}(\lambda=\lambda_1) \times p^{NV^-}_{\pi}(t)
    ,\label{eq:SI_count_dist}
\end{align}
where $\textrm{Poisson}(\lambda)$ is the Poisson distribution with mean $\lambda$ and $p^{NV^{0,-}}_{0,\pi}$ denotes the NV center charge state population for either the neutral (0) or negative (-) charge state with the initial condition of applying a $\pi-$pulse ($\pi$) or not (0). As mentioned above, we assume $\lambda_0=1.6$ and $\lambda_1=6.7$ for the simulation. From the two photon readout distribution, we can then compute the corresponding statistics to extract the readout noise $\sigma_\text{R}$ with \ref{eq:SI_sigmaRgen}
to obtain $\sigma_\text{R}(t_{ion})$, where $t_{ion}$ is the evolution time that corresponds to the duration of the ionization pulse (Extended Data Fig.~\ref{fig:SI_rate_eq}\textbf{a}). The model shows that for a given excitation laser power, as we increase the number of NV centers that we multiplex on (decrease the ionization power per NV center), the optimal ionization pulse time required to obtain the lowest readout noise at a given power increases (Extended Data Fig.~\ref{fig:SI_rate_eq}\textbf{b}), which was also observed qualitatively in experiment. Using the same approach, we investigate the scaling of $\sigma_\text{R}$ as a function of number of multiplexed NV centers for different total available laser power (Extended Data Fig.~\ref{fig:SI_rate_eq}\textbf{c}). With a lower total available laser power, $\sigma_\text{R}$ increases monotonically as the number of multiplexed NV centers increases. With a higher total power, $\sigma_\text{R}$ first decreases and then increases as the number of multiplexed NV centers increases. The scaling can be understood from the physical process behind SCC, where both too high or too low laser power decreases the ionization contrast between the $m_s= 0$ and $m_s=\pm1$ states. We also investigate the effect of the ionization laser wavelength on $\sigma_\text{R}$, which primarily determines the recombination rate from NV$^0$ back to NV$^-$ \cite{aslam_photo-induced_2013}, and obtain a lower floor for $\sigma_\text{R}$ when using a longer wavelength for ionization (Extended Data \cref{fig:SI_rate_eq}\textbf{d}).

\section{Effects of background correlation}
In this section, we model and investigate the effect of background correlations on our measurement. We find that the background correlation acts as a constant shift in the baseline correlation, which allows us to subtract off the baseline from our measured correlation data.

We assume two Poissonian sources with mean value $\mu$: $X_1 \sim \textrm{Poisson}(\mu), X_2 \sim \textrm{Poisson}(\mu)$ that correspond to signals from two NV centers and assume that the signal strength is randomly modulated by a common Gaussian noise source, $N \sim \textrm{Normal}(1,\sigma_N)$. This global noise source models the effect of laser power fluctuations and small drifts of the sample between our SLM coordinate calibrations. For simplicity we assume that both NV centers have the same mean brightness, which holds true for NV centers of the same orientation under the same optical readout power, and define $\sigma_N$ as the strength of the random fluctuation in our detection process. Thus, the detected signal is $S_i=NX_i$, and the correlation between $S_1, S_2$, assuming independent NV centers is 
 
\begin{align}
    \textrm{Corr}(S_1,S_2) = \frac{\mu^2 \sigma_N^2}{\mu^2 \sigma_N^2 + \mu \sigma_N^2 + \mu} \approx \mu \sigma_N^2
    .\label{eq:SI_bg_corr_basecase_result}
\end{align}
The approximation in \ref{eq:SI_bg_corr_basecase_result} holds when $\mu \sigma_N^2, \sigma_N^2 \ll 1$, which is a realistic assumption for our experiment. 

Now consider the case when the the two NV centers are correlated, with a correlation $\textrm{Corr}(X_1, X_2) = r$. The random fluctuation in readout is still assumed to be independent of the NV signal. In this case, we obtain:
\begin{align}
    \textrm{Corr}(S_1,S_2) = \frac{\mu^2 \sigma_N^2 + \mu r (\sigma_N^2 + 1)}{\mu^2 \sigma_N^2 + \mu \sigma_N^2 + \mu} \approx \mu \sigma_N^2 + r(\sigma_N^2 + 1)
    .\label{eq:SI_bg_corr_pos_r_result}
\end{align}

The difference between the measured correlation, $\textrm{Corr}(S_1,S_2)$, and the real correlation between the emitters, $r$ can be computed straightforwardly:
\begin{align}
    \Delta r = \textrm{Corr}(S_1,S_2) - r = \frac{(1-r) \mu^2 \sigma_N^2}{\mu^2 \sigma_N^2 + \mu \sigma_N^2 + \mu} \approx (1-r) \mu \sigma_N^2 \geq 0
    .\label{eq:SI_bg_corr_pos_r_delta}
\end{align}
The approximation in \ref{eq:SI_bg_corr_pos_r_delta} is valid when $\mu \sigma_N^2, \sigma_N^2 \ll 1$. Note that in our case, based on the readout noise we measure, the maximum correlation in NV signal that we should expect from sensing an AC magnetic field signal would be less than 0.5 percent. Therefore, we may simply assume $\Delta r \approx \mu \sigma_N^2$, indicating that the effect of the background correlation is to add the same constant offset to the real correlation between the NV centers no matter whether the two NV centers are correlated or not. The expressions in \ref{eq:SI_bg_corr_pos_r_result} and \ref{eq:SI_bg_corr_pos_r_delta} do not assume a sign for $r$, which means that for two emitters that are either positively or negatively correlated, the effect of the background correlation is adding a constant positive offset regardless of the sign of the correlation. Moreover, a more involved analysis for arbitrary initial photon distributions with mean $\mu$ and variance $\sigma^2$ yields the same conclusion as the above analysis.

We measure the background correlation independently of the correlation experiment that we run to obtain an independent baseline. We read out 15 NV centers simultaneously for 300,000 iterations and compute the resulting pair-wise correlations, shown in Extended Data \cref{fig:SI_bg_corr}. The mean of the measured background correlation is $2.01\times10^{-3}$, which we use as our baseline that we subtract off from the measured raw correlation to give the data in \cref{fig:multiplex_corr}\textbf{c}, \textbf{d}.

\section{Characterization of the NV centers for multiplexed covariance magnetometry}

The magnitude of the measured correlation between NV$_i$ and NV$_j$ is given by 
\begin{align}
    r_{ij}  = \frac{\text{e}^{-[\Tilde{\chi}_i(t_i)+\Tilde{\chi}_j(t_j)]}}{\sigma_{\text{R}_i}\sigma_{\text{R}_j}}\langle \sin[\phi_{C_i}(t_i)] \sin{[\phi_{C_j}(t_j)]} \rangle,\label{eq:SI_correlation}
\end{align}
where $\Tilde{\chi}_i(t)$ is the decoherence function, $\sigma_{\text{R}_i}$ is the spin readout noise, and $\phi_{C_i}$ is the phase accumulated by NV center $i$ due to a correlated field~\cite{rovny_nanoscale_2022}. \ref{eq:SI_correlation} shows that the measured magnetic field correlation is a function of both the NV readout noise and coherence. Since we are working with shallow, near-surface NV centers whose coherence and charge state stability could be corrupted by surface inhomogeneities, we characterize the NV properties before executing the multiplexed covariance experiment. To understand the coherence of the NV centers, we first run a wide-field XY8 dynamical decoupling sequence and sweep the interpulse spacing $\tau$ to characterize the NV coherence time $T_{2,XY8}$. 
The resulting $T_{2,XY8}$ map is shown in Extended Data Fig.~\ref{fig:SI_t2map}. The wide-field $T_{2,XY8}$ data shows that the majority of the measured NV centers have coherence time over 15 \textmu s, indicating that for the interpulse spacing $\tau$\;=\;250\;ns used in our covariance magnetometry sensing sequence, the decoherence factor in correlation is $\sim 0.9^2$. Therefore, in the experiment, we pick NV centers that have coherence time greater than 15 \textmu s.
As for readout noise, the numbers that we measure from the selected NV centers for the multiplexed covariance magnetometry experiments are consistently around $\sigma_\text{R}\approx12$-15 with minor fluctuations possibly due to the regeneration of laser spot pattern.

\newpage

\setcounter{figure}{0}
\renewcommand{\thetable}{\textbf{\arabic{table}}}%

\renewcommand{\figurename}{\textbf{Extended Data Fig.}}

\renewcommand{\thefigure}{\textbf{\arabic{figure}}}%
\clearpage
\section*{Extended Data}

\begin{figure*}[ht]
	\centering
	\includegraphics[width=120mm]{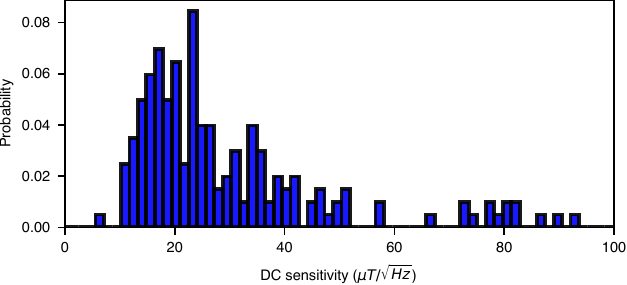}
	\caption{\textbf{Histogram of DC magnetic field sensitivity.} Histogram of the DC magnetic field sensitivity of the NV centers in \cref{fig:spinmeasurements}\textbf{a}. Over 60\% of the NV centers have sensitivities between 10 and 30 \textmu T\;Hz$^{-1/2}$.
    }
	\label{fig:SI_dc_sensitivity}
\end{figure*}

\begin{figure*}[ht]
	\centering
	\includegraphics[width=120mm]{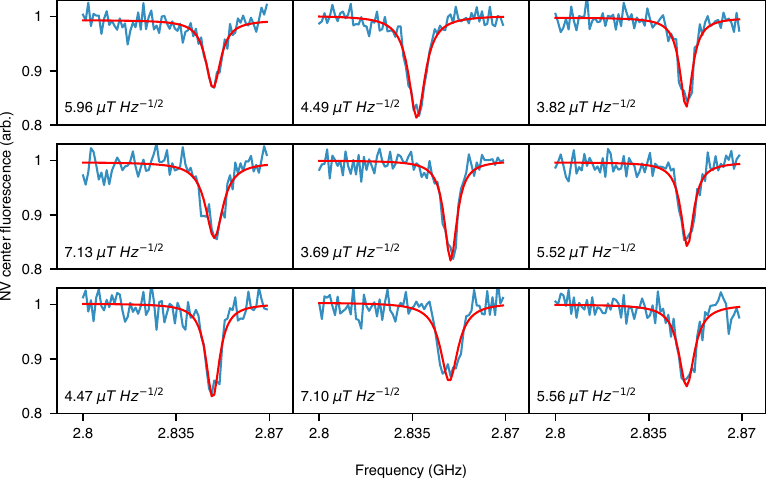}
	\caption{\textbf{DC magnetic field sensitivity measured with a confocal microscope setup} CW-ODMR spectra are measured for 9 NV centers in the same sample using a confocal microscope setup. The corresponding DC magnetic field sensitivities are labeled for each curve.}
	\label{fig:SI_confocal_ODMR}
\end{figure*}

\begin{figure*}[ht]
	\centering
	\includegraphics[width=175mm]{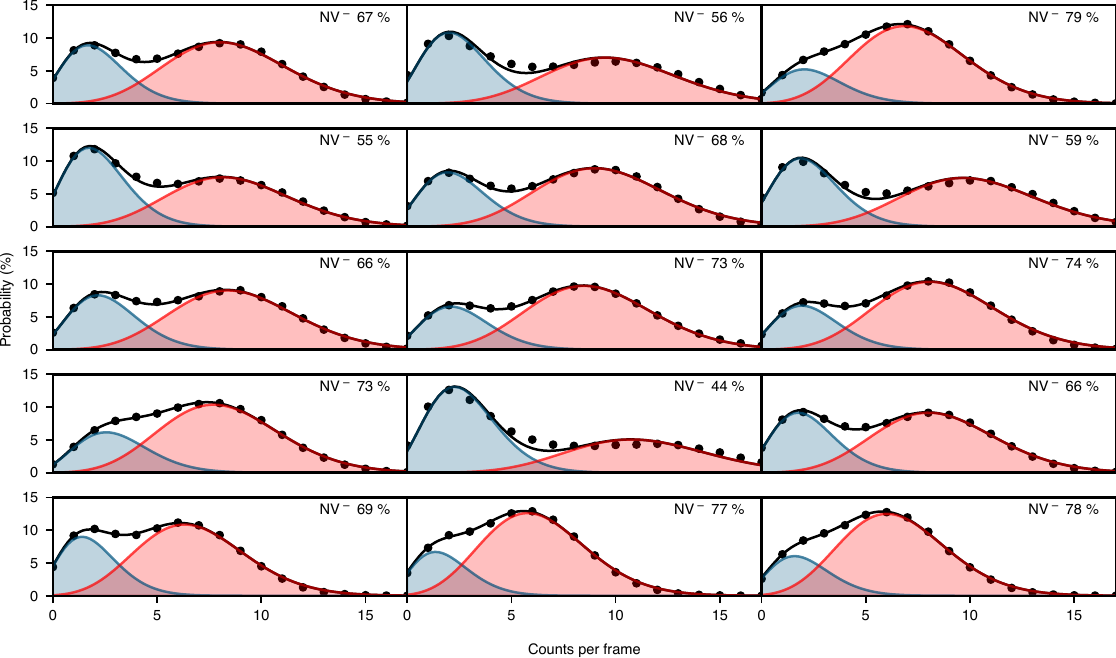}
	\caption{\textbf{Multiplexed charge state readout on 15 NV centers.} 15 NV centers are targeted with the SLM and the charge states are read out simultaneously using the EMCCD. The photon counts per shot of individual NV centers are recorded at each experiment iteration and put into a histogram. The curves are fits to double-Poisson distributions, where the two colors indicate the NV$^0$ (blue) and NV$^-$ (red). The steady state NV$^-$ population of the NV centers is extracted from the relative area under the fit curves of the two Poisson distributions.}
	\label{fig:SI_chargereadout}
\end{figure*}

\begin{figure*}[ht]
	\centering
	\includegraphics[width=89mm]{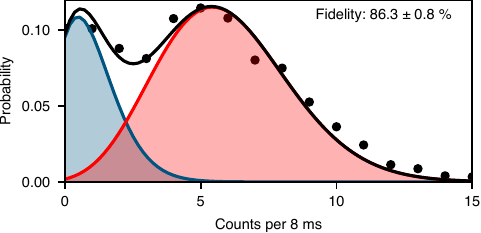}
	\caption{\textbf{Charge state readout using a confocal microscope setup} A typical NV center charge state readout histogram using a confocal microscope is shown. The charge state readout fidelity is 86.3 $\pm$ 0.8 $\%$. The optical setup, the diamond sample, and readout time are the same as the ones in \cref{fig:SLM}\textbf{d}. The curves are double Poisson fit to the histogram.}
	\label{fig:SI_confocal_charge_readout}
\end{figure*}

\begin{figure*}[ht]
	\centering
	\includegraphics[width=120mm]{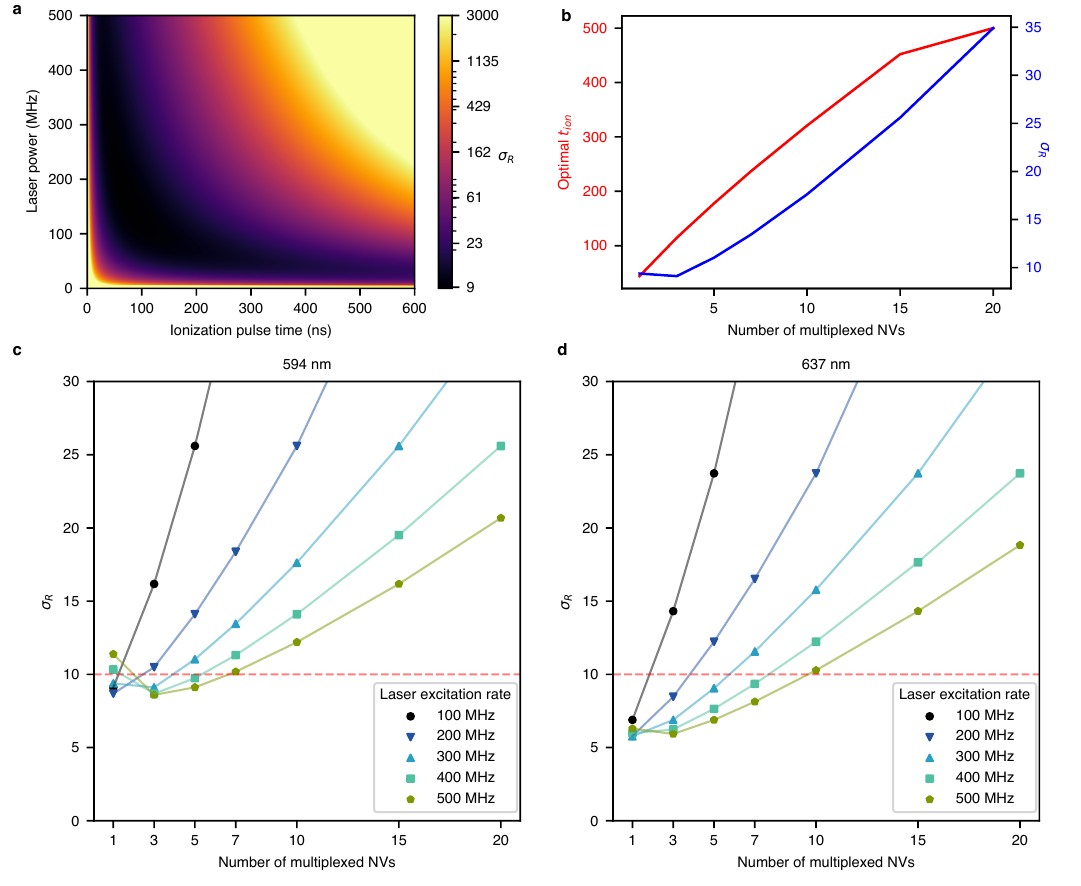}
	\caption{\textbf{Histogram of spin readout noise using multiplexed wide-field green readout.} The spin readout noise of most measured NV centers are well above 100, which is several times worse than that achieved by green readout using a confocal microscope. The increase in readout noise can be attributed to (1) increased background fluorescence for camera readout, (2) lower spin contrast since the camera is always exposing during a pulsed experiment, and (3) low optical power density.
    }
	\label{fig:SI_wf_sigmaR}
\end{figure*}

\begin{figure*}[ht]
	\centering
	\includegraphics[width=183mm]{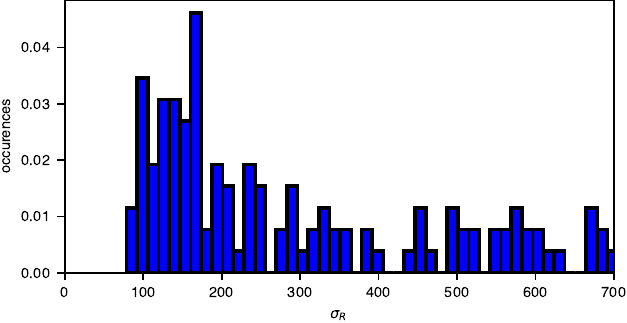}
	\caption{\textbf{Effect of ionization laser power, time, and wavelength on spin state readout noise, $\sigma_\text{R}$}. \textbf{a} $\sigma_\text{R}$ as a function of orange ionization laser power and pulse time simulated using a rate equation model. \textbf{b} $\sigma_\text{R}$ and the optimal ionization pulse duration at a laser power that matches our experiment as a function of number of multiplexed NV centers (inversely proportional to the power per NV center). \textbf{c, d} $\sigma_\text{R}$ as a function of number of multiplexed NV centers for different total laser power for orange ionization laser (\textbf{c}) and red ionization laser (\textbf{d}).}
	\label{fig:SI_rate_eq}
\end{figure*}

\begin{figure*}[ht]
	\centering
	\includegraphics[width=89mm]{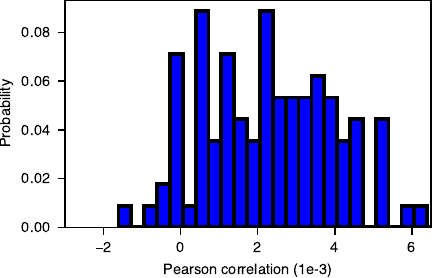}\caption{\textbf{Background correlation data measured on 15 NV centers simultaneously.} We benchmark the magnitude of background correlations by performing multiplexed charge state readout on 15 NV centers 300000 times, then computing all pairwise correlations. The mean of the background correlation distribution is approximately 2e-3, and we subtract this background from all of the covariance magnetometry measurements shown in~\cref{fig:multiplex_corr}.}
	\label{fig:SI_bg_corr}
\end{figure*}

\begin{figure*}[ht]
	\centering
	\includegraphics[width=89mm]{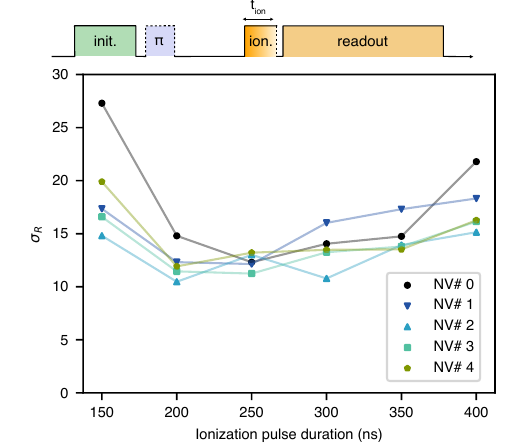}
	\caption{\textbf{Optimizing parameters for multiplexed SCC.} We measure the spin readout noise $\sigma_\text{R}$ of 5 NV centers simultaneously by comparing the readout photon statistics for each NV centers in the $m_s=0$ (no $\pi$-pulse) and $m_s=-1$ ($\pi$-pulse) after the SCC process. To optimize the readout noise as described in the main text, we sweep the ionization pulse time to minimize the readout noise, which is critical for multiplexed covariance magnetometry. In this case, $t_{ion}$ of 250\;ns is considered optimal for all the NV centers.}\label{fig:SI_sigmaR_tion}
\end{figure*}

\begin{figure*}[ht]
	\centering
	\includegraphics[width=89mm]{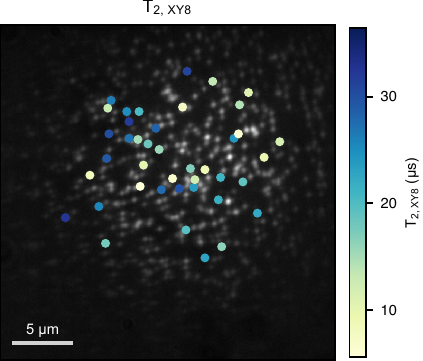}\caption{\textbf{Wide-field spin coherence time map with XY8 dynamical decoupling sequence.} We measure the coherence time $T_{2,XY8}$ for the NV centers of the chosen crystallographic orientation before executing the covariance spectroscopy experiment.}
	\label{fig:SI_t2map}
\end{figure*}

\end{document}